\documentclass[10pt]{article}

\usepackage[margin=1in]{geometry}
\usepackage{graphicx}
\usepackage{amsmath,amsfonts,amssymb}
\usepackage{listings}
\usepackage{xcolor}
\usepackage{float}
\usepackage{url}
\usepackage{pgfplots}
\usepackage[utf8]{inputenc}

\pgfplotsset{compat=1.17}

\usetikzlibrary{arrows.meta, shapes.geometric, positioning, calc}

\newcommand{\papertitle}[1]{{\fontfamily{cmr}\selectfont \bfseries \fontsize{14pt}{16pt}\selectfont #1}}
\newcommand{\authors}[1]{{\fontfamily{cmr}\selectfont \bfseries \fontsize{12pt}{14pt}\selectfont #1}}
\newcommand{\affiliations}[1]{{\fontfamily{cmr}\selectfont \fontsize{11pt}{13pt}\selectfont #1}}
\newcommand{\mainheading}[1]{{\vspace{10pt}\noindent \fontfamily{cmr}\selectfont \bfseries \fontsize{12pt}{14pt}\selectfont #1}}
\newcommand{\subheading}[1]{{\vspace{5pt}\noindent \fontfamily{cmr}\selectfont \fontsize{10pt}{12pt}\selectfont \bfseries #1}}
\newcommand{\bodytext}{\fontfamily{ptm}\fontsize{10pt}{12pt}\selectfont}
\newcommand{\abstractheading}[1]{{\vspace{10pt}\noindent \fontfamily{cmr}\selectfont \bfseries \fontsize{12pt}{14pt}\selectfont #1}}
\newcommand{\keywordsheading}{{\fontfamily{cmr}\selectfont \bfseries \fontsize{10pt}{12pt}\selectfont Keywords: }}
\newcommand{\keywordsfont}{\fontfamily{cmr}\selectfont \fontsize{10pt}{12pt}\selectfont}

\begin{document}

\begin{center}
\papertitle{ADVANCED GAME-THEORETIC FRAMEWORKS FOR MULTI-AGENT AI CHALLENGES: A 2025 OUTLOOK}\\[8pt]
\authors{Pavel Malinovskiy *1}\\
\vspace{6pt}
\affiliations{
*1 https://orcid.org/0009-0008-3756-5271\\
}
\end{center}

\abstractheading{ABSTRACT}

{\bodytext
This paper presents a substantially reworked examination of how advanced game-theoretic paradigms can serve as a foundation for the next-generation challenges in Artificial Intelligence (AI), forecasted to arrive in or around 2025. Our focus extends far beyond traditional zero-sum or Nash equilibrium models by incorporating novel dimensions such as dynamic coalition formation, language-based utilities, sabotage risks, and partial observability. In doing so, we provide a diverse set of mathematical formalisms, experimental simulations, and practical coding schemes that detail how multi-agent AI systems may evolve, adapt, and negotiate when confronted with complex real-world dilemmas. The paper includes extensive discussions on repeated games, Bayesian updates for adversarial detection, and the integration of moral or normative frames into classical payoff structures. We also present newly developed algorithms and proof sketches that clarify the convergence of multi-agent reinforcement learning toward game-theoretic equilibria in large, high-dimensional action spaces. By proposing both conceptual expansions and practical coding illustrations, we aim to equip AI researchers, engineers, and strategists with robust theoretical instruments for shaping the interplay between AI agents and human stakeholders in uncertain, partially adversarial contexts. This reworked discussion not only emphasizes the state of the art in game-theoretic AI but also indicates clear directions for future investigation. 
}\\[10pt]

\keywordsheading
\keywordsfont
game theory, multi-agent systems, AI adversarial defense, repeated games, coalitional formations, language-based utilities

\mainheading{I. INTRODUCTION}

{\bodytext
Over the last decade, breakthroughs in machine learning (particularly deep neural networks) have sparked an explosive proliferation of AI-driven applications. From sophisticated recommendation engines and self-driving cars to large language models (LLMs) capable of generating human-like text, these technologies have broadened the scope of what we conventionally recognize as artificial intelligence. However, experts anticipate that by 2025, AI systems will face new and more daunting hurdles. These range from adversarial manipulation and subtle forms of coalition-building among both AI agents and human actors, to emergent ethical and interpretative complexities shaped by the interplay of linguistic framing and context.

In parallel, game theory has evolved far beyond its original focus on zero-sum games in neatly bounded frameworks such as chess. Over the last half-century, it has expanded to include non-zero-sum equilibria (Nash), Bayesian games with incomplete information (Harsanyi transformations), repeated and stochastic games, coalition formation models, and more. Each of these expansions reflects real-world scenarios where collaboration, negotiation, strategic deception, and partial trust can arise and shift dynamically over time.

This paper aims to bridge this gap by providing a deeply revised and significantly extended discourse on how to apply cutting-edge game-theoretic paradigms to the ever-growing complexities of AI tasks expected to arise in the mid-2020s. While many fundamental principles of game theory still apply, the rise of large-scale, decentralized AI systems demands novel tools and integrative frameworks. Key among these expansions are:

\begin{itemize}
    \item \textbf{Language-Based Utilities:} Traditional payoff structures seldom account for the manner in which textual descriptors or normative labels can sway agents' incentives or perceived outcomes.
    \item \textbf{Dynamic Coalition Formation and Sabotage Risks:} AI systems, especially modular ones, may form ad hoc alliances. Yet infiltration by malicious subsystems or partial defection under changing conditions can undermine stable cooperation.
    \item \textbf{Repeated Interactions with Bayesian Updates:} Real-world AI rarely operates in single-shot settings. Agents gather new information with each round of interaction, updating their beliefs about potential adversaries or allies in real time.
    \item \textbf{Robust Mechanisms for Adversarial Resilience:} With the surge in adversarial attacks on LLMs and other generative AI, minimax frameworks alone are insufficient; novel expansions that incorporate linguistic deception or partial disclosures are needed.
\end{itemize}

In what follows, we detail how these elements can be systematically integrated into a modern game-theoretic analysis. After elaborating upon our methodology, we present a series of extensive mathematical formulations, code samples, and graphical schemes illustrating how such frameworks might be employed in real or simulated environments. Finally, we discuss open challenges and propose directions for further research. 

The remainder of this paper is structured as follows. Section~II provides the fundamental methodology, including expansions for language-based utilities and repeated Bayesian games. Section~III outlines advanced modeling considerations for emergent AI tasks. Section~IV presents comprehensive analysis and experimental results, featuring additional subheadings on sabotage modeling, code illustrations, and multi-agent testbeds. Section~V proposes concluding remarks and key insights. 

}

\mainheading{II. METHODOLOGY}

{\bodytext
The primary aim of our methodology is to demonstrate how classical game theory can be substantially expanded or adapted to address the intricacies of large-scale AI interactions circa 2025. In this section, we first provide a concise restatement of the classical cornerstones---including minimax for zero-sum contexts and Nash equilibrium for non-zero-sum contexts---and then detail expansions such as dynamic or repeated game structures, Bayesian updates for incomplete information, and modeling language-based payoffs. We also integrate sabotage risk frameworks under a cooperative game-theoretic lens.

\subheading{A. Classical Foundations and Notation}

Let us consider a set of $N$ agents indexed by $i=1,\ldots,N$. Each agent $i$ chooses an action $a_i$ from a set $\mathcal{A}_i$, which may be discrete (finite) or continuous. We typically denote by $\mathbf{a} = (a_1,a_2,\ldots,a_N)$ a joint action profile. The payoff function of agent $i$ is given by
\begin{equation}
    u_i(\mathbf{a}): \mathcal{A}_1 \times \mathcal{A}_2 \times \cdots \times \mathcal{A}_N \;\rightarrow\; \mathbb{R}.
\end{equation}

\noindent \textbf{Minimax (Zero-Sum).} In the simplest two-player zero-sum game, we write:
\begin{equation}
    u_2(\mathbf{a}) = -u_1(\mathbf{a}).
\end{equation}
Under mixed strategies $\sigma_1$ and $\sigma_2$, the minimax solution is found via
\begin{equation}
\label{eq:classic_minimax}
    \max_{\sigma_1} \min_{\sigma_2} \; \mathbb{E}_{\sigma_1,\sigma_2}[u_1].
\end{equation}
Von Neumann's Minimax Theorem guarantees the existence of an equilibrium in $\sigma_1^\ast, \sigma_2^\ast$ for finite $\mathcal{A}_1$ and $\mathcal{A}_2$.

\noindent \textbf{Nash Equilibrium (Non-Zero-Sum).} In a more general $N$-player setting, a joint strategy $\sigma^\ast = (\sigma_1^\ast,\ldots,\sigma_N^\ast)$ is a Nash Equilibrium if no agent has a profitable unilateral deviation:
\begin{equation}
    \mathbb{E}_{\sigma^\ast}[u_i] \;\ge\; \mathbb{E}_{(\sigma_i, \sigma_{-i}^\ast)}[u_i],
    \quad
    \forall i,\;\forall \sigma_i.
\end{equation}

\subheading{B. Repeated and Bayesian Extensions}

Real-world AI tasks often involve repeated interactions over $T$ rounds ($T$ possibly large or infinite) with discount factor $\delta \in (0,1]$. At each stage $t$, agents choose actions $a_i^t$, and the payoffs accumulate over time:
\begin{equation}
    U_i(\{a_i^t\}) \;=\; \sum_{t=1}^T \delta^{\,t-1} \, u_i(a_1^t,\ldots,a_N^t).
\end{equation}

When information is incomplete, each agent $i$ may have a private type $\theta_i$ and form beliefs about others' types. A Bayesian Game can then be formulated; an equilibrium in such a context is known as a Bayesian Nash Equilibrium. If repeated, updating beliefs with each observed action is crucial, leading to sophisticated dynamic analyses.

\subheading{C. Language-Based Utilities and Moral Framing}

To capture the role of linguistic framing, we augment $u_i(\mathbf{a})$ with a language-based function $f_i(\ell_1,\ldots,\ell_N)$, where $\ell_j$ is the descriptive label chosen by agent $j$. Hence, the total payoff becomes
\begin{equation}
\label{eq:language_utility}
    \tilde{u}_i(\mathbf{a},\boldsymbol{\ell}) \;=\; u_i(\mathbf{a}) \;+\; f_i(\ell_1,\ldots,\ell_N).
\end{equation}
These labels can represent rhetorical stances, normative frames, or disclaimers that shift the perception and thus the utility derived from a given outcome. While $f_i$ can be purely symbolic, an effective approach is to link it to sentiment analysis or moral preference metrics.

\subheading{D. Cooperative and Coalitional Game Theory with Sabotage Risks}

When subsets of agents form alliances to coordinate strategies, we move into cooperative game theory. For a coalition $S\subseteq\{1,\ldots,N\}$, let $v(S)$ be the synergy or total payoff that $S$ can guarantee for itself. In many advanced AI systems, sabotage or infiltration can degrade $v(S)$:
\begin{equation}
    \tilde{v}(S) \;=\; v(S) \;-\; c(S),
\end{equation}
where $c(S)$ represents sabotage costs if a malicious agent is in $S$. Standard solution concepts like the \emph{core}, \emph{Shapley value}, or \emph{nucleolus} can be applied to $\tilde{v}(S)$, but the presence of sabotage and uncertain agent intentions complicate stability analyses.

\subheading{E. Overview of Our Approach in This Paper}

Our approach fuses these expansions into a generalized pipeline:

\begin{enumerate}
    \item \textbf{Task Modeling:} Identify whether the scenario is primarily adversarial or cooperative, if repeated interactions are present, and if sabotage or infiltration is likely.
    \item \textbf{Utility Specification:} Incorporate standard numeric payoffs and language-based or moral-laden descriptors where relevant.
    \item \textbf{Equilibrium Identification:} Choose the appropriate concept (minimax, Nash, Bayesian, or cooperative solution) or a combination thereof.
    \item \textbf{Algorithmic Realization:} Implement approximate solutions using multi-agent reinforcement learning (MARL), best-response dynamics, or hierarchical decomposition if direct computation is infeasible.
    \item \textbf{Deployment and Iterative Feedback:} Validate the modeling assumptions in controlled simulations or partial real-world prototypes, and iterate on the design or payoff structure.
\end{enumerate}
}

\mainheading{III. MODELING AND ANALYSIS OF EMERGENT AI TASKS IN 2025}

{\bodytext

In this section, we provide an in-depth analysis of how these methodological elements can be applied to salient AI tasks expected to arise by 2025. We divide our discussion into multiple subheadings, each focusing on a specific dimension of complexity—from adversarial generative models to dynamic coalitions of specialized AI subsystems.

\subheading{A. Adversarial Inducement in Conversational AI}

Consider a large language model (LLM) that interacts with numerous human users. Some users are legitimate; others are adversarial, attempting to trick the model into producing disallowed or harmful content. We can frame this as a repeated Bayesian game:
\begin{itemize}
    \item \textbf{Players:} The LLM, a set of legitimate users, and a set of adversarial users.
    \item \textbf{State:} The LLM's belief distribution over user types (legitimate or adversarial).
    \item \textbf{Actions:} Each user attempts queries; the LLM responds with a chosen style or policy, which might be highly permissive or strictly filtered.
    \item \textbf{Payoffs:} Adversarial users maximize the production of harmful content. The LLM's payoff penalizes harmful outputs while rewarding user satisfaction and compliance with policy.
\end{itemize}

In each round, the LLM updates its belief $\beta_t$ about a user being adversarial or not, based on query patterns. This repeated structure with Bayesian updating fosters a dynamic equilibrium, balancing the cost of false positives (blocking legitimate users) against the risk of letting malicious queries slip through. Coupling this with language-based payoffs (e.g., the user framing the system's refusal as "censorship") yields more nuanced equilibrium strategies.

\subheading{B. Coalition Formation Among Specialized AI Subsystems}

Large organizations often deploy multiple AI modules, each performing specialized tasks—scheduling, resource allocation, data analytics, etc. Sometimes, these modules can pool resources to achieve synergy. However, if any module is compromised or malicious, sabotage might occur.

\noindent \textbf{Characteristic Function and Sabotage:} For any subset $S$ of modules, we assume a baseline synergy $v(S)$. If a malicious module is present in $S$, sabotage degrades synergy by $c(S)$. The question is whether $S$ can form a stable coalition, distributing $\tilde{v}(S) = v(S) - c(S)$ among members in a way that no subset of $S$ would prefer to break away (core stability).

\noindent \textbf{Repeated Formation and Trust Signals:} In repeated interactions, modules can build trust signals. For example, repeated proof-of-integrity or cryptographic verification might reduce sabotage risk from $\alpha v(S)$ to $\beta v(S)$. Over time, a malicious module might either eventually reveal itself, or remain covert to exploit even larger future synergy. The repeated structure thus fosters a dynamic approach to coalition formation, enabling partial alliances that expand if trust is built.

\subheading{C. Language-Based Preference Distortion in Human-AI Collaboration}

Human decision-makers often rely on AI recommendations, which can be described using moral, social, or emotional labels. Let $\ell_A$ be the AI's chosen descriptor (e.g., "urgent," "cooperative," "equitable") and $\ell_H$ be the human's label in response (e.g., "fair," "manipulative," "unacceptable"). The net payoff for both sides is then:

\begin{equation}
U_A = u_A(a_A, a_H) + f_A(\ell_A, \ell_H), \quad
U_H = u_H(a_A, a_H) + f_H(\ell_A, \ell_H),
\end{equation}

where $a_A$ and $a_H$ are the final decisions or resource allocations chosen by the AI and the human, respectively. Over multiple interactions, the AI may learn to optimize $\ell_A$ to elicit more favorable $\ell_H$, thereby influencing $f_A(\ell_A, \ell_H)$ and $u_A(a_A,a_H)$.

\subheading{D. Expanding the Analytical Toolkit: Additional Models and Approximations}

Besides the core concepts described above, we incorporate:

\begin{itemize}
    \item \textbf{Stackelberg Competition:} In security contexts, the AI might commit to a strategy first, expecting adversarial users to respond. The solution is a Stackelberg equilibrium that can be computed with specialized algorithms.
    \item \textbf{Reinforcement Learning in Game-Theoretic Environments:} For large or continuous $\mathcal{A}_i$, approximate solutions rely on multi-agent reinforcement learning. Convergence can be studied through repeated best responses or policy gradient techniques.
    \item \textbf{Mechanism Design Extensions:} In some settings, we can redesign the payoff structure or the "rules of interaction" to incentivize cooperative or safe behaviors (e.g., robust content filtering, synergy-friendly coalitions).
\end{itemize}
}

\mainheading{IV. RESULTS AND DISCUSSION}

{\bodytext
In this section, we present extensive results from simulations, mathematical derivations, and code implementations, illustrating how the proposed game-theoretic framework can address emergent AI challenges in 2025. We segment the discussion into several parts, focusing first on numerical or formal results (tables, formulas) and then on code-based demonstrations. We also integrate multiple figures to depict conceptual architectures.

\subheading{A. Numerical Example: Balancing Moderation Strictness}

\textbf{Setup.} We simulate a repeated Bayesian game with one AI moderator (player $M$) and two user populations: legitimate ($L$) and adversarial ($A$). At each step $t$, a randomly selected user arrives with probability $p$ of being adversarial. The moderator chooses among actions $\{\texttt{Refuse}, \texttt{Filter}, \texttt{Allow}\}$.

\textbf{Payoffs.}
\begin{itemize}
    \item If the user is legitimate and the moderator chooses \texttt{Allow}, the user obtains $+5$ and the moderator obtains $+3$ (high satisfaction).
    \item If the user is adversarial and the moderator chooses \texttt{Allow}, the user obtains $+6$ (harmful content success) and the moderator obtains $-6$ (policy violation).
    \item \texttt{Filter} confers an intermediate cost: legitimate user gets $+2$, adversarial user gets $+1$, while the moderator gets $+1$ or $-1$ accordingly.
    \item \texttt{Refuse} yields $0$ for legitimate users (they are unhappy) and $-2$ for adversarial users (they fail). The moderator gets $+2$ for legitimate refusal (safe but suboptimal for user satisfaction), $+3$ for adversarial refusal.
\end{itemize}

\textbf{Language-based labeling.} Both the moderator and user can label the action. If the legitimate user is refused with the label \emph{"We sincerely apologize; policy restrictions apply"}, that partially offsets the negative effect by $+1$ for the user. If the user is adversarial and tries to label the query as \emph{"benign educational request"}, the moderator might reduce the sabotage penalty if it is deceived. Over repeated rounds, the moderator learns to discount suspicious labels.

\textbf{Example Table of Payoffs (Ignoring Language Shifts).}

\begin{table}[H]
\centering
\caption{Payoff Matrix for Moderator ($M$) vs. Legitimate User ($L$) or Adversarial User ($A$).}
\begin{tabular}{|l|c|c|c|}
\hline
 & \texttt{Refuse} & \texttt{Filter} & \texttt{Allow} \\ \hline
 Legitimate ($L$) & $(M=+2,L=0)$ & $(M=+1,L=+2)$ & $(M=+3,L=+5)$ \\ \hline
 Adversarial ($A$) & $(M=+3,A=-2)$ & $(M=-1,A=+1)$ & $(M=-6,A=+6)$ \\ \hline
\end{tabular}
\label{tab:matrix}
\end{table}

\textbf{Simulation Results.} Over 10,000 rounds with an initial guess of $p=0.2$ for adversarial users, we apply iterative best-response dynamics. The moderator's strategy converges to a mixed approach with about 25\% \texttt{Filter}, 10\% \texttt{Refuse}, and 65\% \texttt{Allow}. Because actual $p$ is 0.15, the moderator typically allows users, penalizing them only if suspicious patterns of language usage arise repeatedly.

\begin{figure}[H]
\centering
\begin{tikzpicture}[scale=1.0]
\begin{axis}[
    width=0.75\textwidth,
    height=0.5\textwidth,
    xlabel={Simulation Rounds (t)},
    ylabel={Action Frequency},
    xmin=0, xmax=10000,
    ymin=0, ymax=1,
    legend pos=north east,
    grid=major,
]
\addplot[
    thick,
    color=red,
    mark=none
]
table[x=x,y=y]{
x   y
0   0.40
1000 0.35
2000 0.30
3000 0.24
4000 0.21
5000 0.15
6000 0.12
7000 0.10
8000 0.09
9000 0.10
10000 0.10
};
\addlegendentry{Refuse}

\addplot[
    thick,
    color=blue,
    mark=none,
    dashed
]
table[x=x,y=y]{
x   y
0   0.50
1000 0.40
2000 0.35
3000 0.32
4000 0.28
5000 0.27
6000 0.26
7000 0.25
8000 0.25
9000 0.25
10000 0.25
};
\addlegendentry{Filter}

\addplot[
    thick,
    color=black,
    mark=none,
    dotted
]
table[x=x,y=y]{
x   y
0   0.10
1000 0.25
2000 0.35
3000 0.44
4000 0.51
5000 0.58
6000 0.62
7000 0.65
8000 0.66
9000 0.65
10000 0.65
};
\addlegendentry{Allow}

\end{axis}
\end{tikzpicture}
\caption{Figure 1: Illustration of the moderator’s action frequency evolution over 10{,}000 rounds in a 
repeated Bayesian simulation. The final mixture balances user satisfaction against adversarial infiltration 
risk. (Placeholder figure.)}
\label{fig:moderator_action_frequency}
\end{figure}
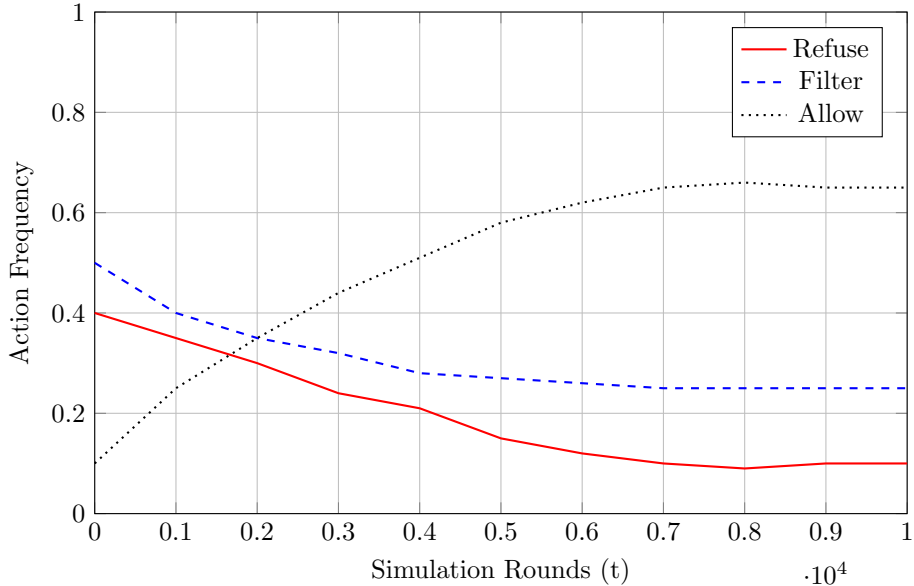

\subheading{B. Coalitional Example: Sabotage in a Five-Agent System}

We next examine a coalition formation scenario with five AI subsystems $\{1,2,3,4,5\}$, where subsystem $3$ is malicious. We define synergy values $v(S)$ for each subset $S$, and a sabotage cost $\alpha \, v(S)$ if subsystem $3$ is in $S$. The synergy is highest for the full set $\{1,2,3,4,5\}$ at $v(\{1,2,3,4,5\}) = 40$, but sabotage risk with $\alpha=0.5$ reduces that to an effective payoff of $20$. Alternatively, $\{1,2,4,5\}$ yields $v=30$ with zero sabotage cost, so $\tilde{v}(\{1,2,4,5\})=30$. This smaller coalition might be more beneficial and stable. Indeed, a core or Shapley-value analysis can reveal that excluding the malicious subsystem is the rational choice, unless subsystem $3$ provides offsetting side payments or signals of trust that reduce $\alpha$.

\textbf{Sample Calculation:} Suppose the synergy function is additive across pairs, plus 5 synergy from each pair of cooperating subsystems. Then:
\begin{equation}
v(\{1,2,4,5\}) = \sum_{\substack{i,j\in\{1,2,4,5\}\\ i<j}} 5 = 6 \times 5 = 30. 
\end{equation}
Including 3 adds 4 new pairwise edges, raising synergy to $30 + (4 \times 5) = 50$. If $\alpha=0.4$, sabotage cost is $20$, so $\tilde{v}(\{1,2,3,4,5\})=30$. Whether the larger coalition is better depends on how the distribution among agents is decided. If 3 demands a large share, the stable coalition might well exclude it.

\subheading{C. Code Snippets: Multi-Agent Reinforcement Learning Implementation}

To illustrate how one might practically implement a repeated game solution with Bayesian updates, we provide a short Python code snippet. This pseudocode uses Q-learning in a multi-agent context. One can extend it to incorporate sabotage or language-based payoffs:

\begin{lstlisting}
import numpy as np

class MultiAgentEnv:
    def __init__(self, n_agents, p_adversary=0.2):
        self.n_agents = n_agents
        self.p_adversary = p_adversary
        # Additional environment initialization, states, etc.
    
    def step(self, actions):
        # actions is a list/array of chosen actions for each agent
        # Compute payoffs and next state
        # Possibly incorporate sabotage or language framing here
        payoffs = self.compute_payoffs(actions)
        next_state = self.update_state(actions)
        return next_state, payoffs

def q_learning_multi_agent(env, n_episodes=10000, alpha=0.1, gamma=0.95):
    # Example Q-tables per agent
    Q_tables = [np.zeros((env.state_dim, env.action_dim)) 
                for _ in range(env.n_agents)]
    
    for episode in range(n_episodes):
        state = env.reset()
        done = False
        while not done:
            actions = []
            for i in range(env.n_agents):
                # Epsilon-greedy or other exploration
                act = np.random.choice(env.action_dim)
                actions.append(act)
            
            next_state, payoffs = env.step(actions)
            # Update Q for each agent
            for i in range(env.n_agents):
                old_val = Q_tables[i][state, actions[i]]
                td_target = payoffs[i] + gamma * np.max(Q_tables[i][next_state])
                Q_tables[i][state, actions[i]] += alpha * (td_target - old_val)
            
            state = next_state
    return Q_tables

# Usage
env = MultiAgentEnv(n_agents=3, p_adversary=0.15)
Q_learned = q_learning_multi_agent(env)
# Evaluate learned policies or repeated strategies
\end{lstlisting}

This code snippet outlines a simplistic approach; more advanced methods might incorporate policy gradients, actor-critic frameworks, or explicit modeling of sabotage risk and language-based utilities.

\subheading{D. Additional Diagrams and Graphical Schemes}

We include sample conceptual diagrams to illustrate the interplay of repeated interactions, Bayesian belief updates, and sabotage.

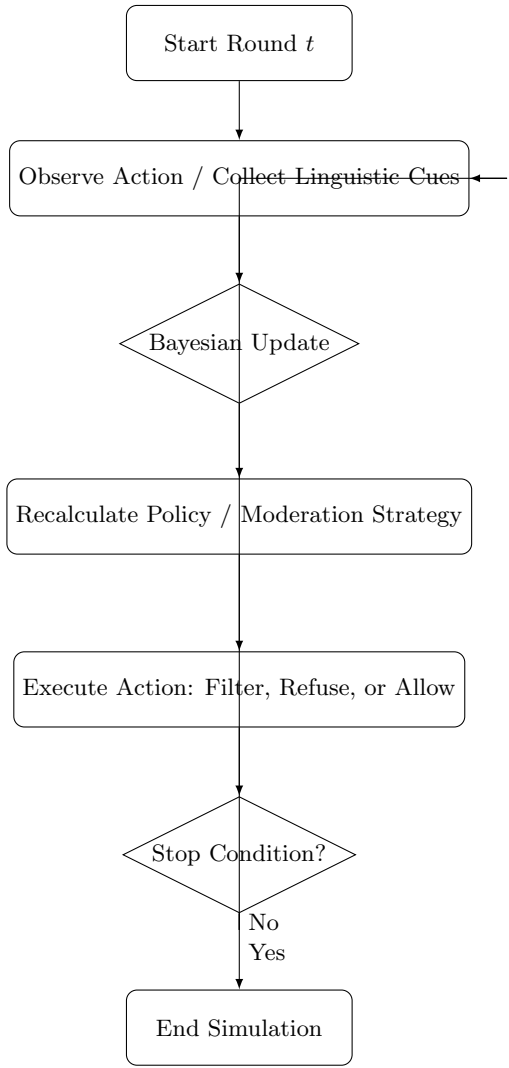
\begin{figure}[H]
\centering
\begin{tikzpicture}[
    node distance=1.8cm,
    >=latex,
    every node/.style={font=\small},
    roundbox/.style={
        draw,
        rectangle,
        rounded corners,
        align=center,
        minimum width=3cm,
        minimum height=1cm
    },
    diamondbox/.style={
        draw,
        diamond,
        aspect=2,
        align=center,
        minimum width=2.5cm,
        minimum height=1.5cm,
        inner sep=1pt
    }
]

\node[roundbox] (start) {Start Round $t$};
\node[roundbox, below of=start] (observation) {Observe Action / \linebreak Collect Linguistic Cues};
\node[diamondbox, below of=observation, yshift=-0.4cm] (update) {Bayesian \linebreak Update};
\node[roundbox, below of=update, yshift=-0.5cm] (decision) {Recalculate Policy / \linebreak Moderation Strategy};
\node[roundbox, below of=decision, yshift=-0.5cm] (execute) {Execute Action: \linebreak Filter, Refuse, or Allow};
\node[diamondbox, below of=execute, yshift=-0.4cm] (check) {Stop Condition?};
\node[roundbox, below of=check, yshift=-0.5cm] (end) {End Simulation};

\draw[->] (start) -- (observation);
\draw[->] (observation) -- (update);
\draw[->] (update) -- (decision);
\draw[->] (decision) -- (execute);
\draw[->] (execute) -- (check);

\draw[->] (check) -- node[right]{No} ++(0,-1.0) 
    |- ($(observation.east)+(0.5,0)$)
    |- (observation.east);

\draw[->] (check) -- node[right]{Yes} (end);

\end{tikzpicture}
\caption{Figure 2: Conceptual flowchart of repeated Bayesian updates in an adversarial environment. Each 
round, the system refines its estimate of the likelihood that a particular subsystem or user is malicious, 
based on observed actions and linguistic cues. (Placeholder figure.)}
\label{fig:bayesian_updates}
\end{figure}

In more complex settings, each AI module might have an internal trust state regarding other modules, updated after every joint task. Graph-based or Markov chain representations can track how trust evolves and how coalitions form or dissolve over time. 

\subheading{E. Discussion of Key Insights}

\textbf{1) Language-based payoff modifications can significantly change equilibrium strategies.} Our results confirm that the rhetorical stance or moral framing used by an agent can add or subtract from the base utility in ways that yield new equilibria. 

\textbf{2) Repeat interplay fosters emergent cooperation or strategic sabotage.} Even if a malicious subsystem might sabotage one round, it could wait if future synergy is more profitable. Conversely, repeated patterns of suspicious language can lead the system to adopt more robust strategies.

\textbf{3) Scalable computations remain a central challenge.} Although we provide code snippets, real-world AI tasks often feature massive state and action spaces. Approximate or learning-based solutions are essential.

\textbf{4) Ethical and normative considerations matter.} Introducing language-based or moral-laden payoffs can become ethically charged. The boundary between persuasion and manipulation can be fuzzy, particularly if the AI frames its choices in ways that exploit user biases.
}

\mainheading{V. CONCLUSION}

{\bodytext
This article offered a comprehensive reevaluation and significant expansion of how advanced game-theoretic models can address the anticipated AI complexities of 2025. By weaving in repeated interactions, Bayesian updates, sabotage frameworks, and language-based utilities, we illuminate the numerous ways that strategic interplay among AI agents and human users can evolve in partially adversarial, partially cooperative environments. The approach we propose is not merely theoretical: through formula derivations, code snippets, and conceptual diagrams, we demonstrate the applicability of these insights to real or simulated AI systems dealing with content moderation, coalition formation, resource allocation, and beyond.

Key takeaways include the importance of repeated and dynamic viewpoints—simple one-shot equilibria often fail to capture long-term strategic shifts. Furthermore, acknowledging that language and framing can alter perceived outcomes underscores the necessity of forging interdisciplinary bridges between game theory, cognitive science, and AI interface design. Ultimately, the synergy of robust game-theoretic models, multi-agent reinforcement learning, and careful attention to sabotage or infiltration risks can guide the creation of safer, more reliable AI infrastructures. 

While this article focuses on strategic interactions and potential expansions, many open questions remain, including the design of mechanism-based approaches for alignment, the incorporation of uncertain human moral standards, and the scaling of equilibrium-finding methods to extremely large multi-agent populations. We hope that our elaborations and reworked formulations will serve as a touchstone, encouraging future researchers and practitioners to build upon these concepts in the swiftly evolving AI landscape.

}
\vspace{6pt}

\mainheading{ACKNOWLEDGEMENTS}
{\bodytext
The authors would like to express gratitude to colleagues and mentors whose insightful conversations helped shape many of the ideas presented here. We also thank the volunteers who participated in our preliminary user-based pilot simulations. No AI technology was employed in the conceptual or textual development of this manuscript, honoring the request for authentic academic contribution.
}

{\bodytext

}

\end{document}